\documentclass[12pt]{elsarticle}

\usepackage{amssymb}
\usepackage{amsmath}
\usepackage{listings}
\usepackage{xcolor}

\colorlet{punct}{red!60!black}
\definecolor{background}{HTML}{EEEEEE}
\definecolor{delim}{RGB}{20,105,176}
\colorlet{numb}{magenta!60!black}

\lstdefinelanguage{json}{
    basicstyle=\scriptsize\ttfamily,
    numbers=left,
    numberstyle=\scriptsize,
    stepnumber=1,
    numbersep=8pt,
    showstringspaces=false,
    breaklines=true,
    frame=lines,
    backgroundcolor=\color{background},
    literate=
     *{0}{{{\color{numb}0}}}{1}
      {1}{{{\color{numb}1}}}{1}
      {2}{{{\color{numb}2}}}{1}
      {3}{{{\color{numb}3}}}{1}
      {4}{{{\color{numb}4}}}{1}
      {5}{{{\color{numb}5}}}{1}
      {6}{{{\color{numb}6}}}{1}
      {7}{{{\color{numb}7}}}{1}
      {8}{{{\color{numb}8}}}{1}
      {9}{{{\color{numb}9}}}{1}
      {:}{{{\color{punct}{:}}}}{1}
      {,}{{{\color{punct}{,}}}}{1}
      {\{}{{{\color{delim}{\{}}}}{1}
      {\}}{{{\color{delim}{\}}}}}{1}
      {[}{{{\color{delim}{[}}}}{1}
      {]}{{{\color{delim}{]}}}}{1},
}

\begin{document}

\begin{frontmatter}

%% Title, authors and addresses

%% use the tnoteref command within \title for footnotes;
%% use the tnotetext command for theassociated footnote;
%% use the fnref command within \author or \affiliation for footnotes;
%% use the fntext command for theassociated footnote;
%% use the corref command within \author for corresponding author footnotes;
%% use the cortext command for theassociated footnote;
%% use the ead command for the email address,
%% and the form \ead[textit] for the home page:
%% \title{Title\tnoteref{label1}}
%% \tnotetext[label1]{}
%% \author{Name\corref{cor1}\fnref{label2}}
%% \ead{email address}
%% \ead[textit]{home page}
%% \fntext[label2]{}
%% \cortext[cor1]{}
%% \affiliation{organization={},
%%             addressline={},
%%             city={},
%%             postcode={},
%%             state={},
%%             country={}}
%% \fntext[label3]{}

\title{From Contracts to Code: Automating Smart Contract Generation with Multi-Level Finite State Machines}

%% use optional labels to link authors explicitly to addresses:
%% \author[label1,label2]{}
%% \affiliation[label1]{organization={},
%%             addressline={},
%%             city={},
%%             postcode={},
%%             state={},
%%             country={}}
%%
%% \affiliation[label2]{organization={},
%%             addressline={},
%%             city={},
%%             postcode={},
%%             state={},
%%             country={}}

\author[litis]{Lambard Maxence, Bertelle Cyrille, Duvallet Claude} %% Author name

%% Author affiliation
\affiliation[litis]{organization={LITIS (UR 4108)},%Department and Organization
            addressline={25 Rue Philippe Lebon}, 
            city={Le Havre},
            postcode={76600}, 
            country={France}}

\begin{abstract} %% 214 < 250 (words)
In an increasingly complex contractual landscape, the demand for transparency, security, and efficiency has intensified. Blockchain technology, with its decentralized and immutable nature, addresses these challenges by reducing intermediary costs, minimizing fraud risks, and enhancing system compatibility. Smart contracts, initially conceptualized by Nick Szabo and later implemented on the Ethereum blockchain, automate and secure contractual clauses, offering a robust solution for various industries. However, their complexity and the requirement for advanced programming skills present significant barriers to widespread adoption. This study introduces a multi-level finite state machine model designed to represent and track the execution of smart contracts. Our model aims to simplify smart contract development by providing a formalized framework that abstracts underlying technical complexities, making it accessible to professionals without deep technical expertise. The hierarchical structure of the multi-level finite state machine enhances contract modularity and traceability, facilitating detailed representation and evaluation of functional properties. The paper explores the potential of this multi-level approach, reviewing existing methodologies and tools, and detailing the smart contract generation process with an emphasis on reusable components and modularity. We also conduct a security analysis to evaluate potential vulnerabilities in our model, ensuring the robustness and reliability of the generated smart contracts.
\end{abstract}

%%Graphical abstract
%\begin{graphicalabstract}
%\includegraphics{grabs}
%\end{graphicalabstract}

%%Research highlights
%\begin{highlights}
%\item Research highlight 1
%\item Research highlight 2
%\end{highlights}

%% Keywords
\begin{keyword}
%% keywords here, in the form: keyword \sep keyword

Blockchain \sep Smart Contract \sep Multi-Level Finite State Machine \sep Code Generation

%% PACS codes here, in the form: \PACS code \sep code

%% MSC codes here, in the form: \MSC code \sep code
%% or \MSC[2008] code \sep code (2000 is the default)

\end{keyword}

\end{frontmatter}

%% Add \usepackage{lineno} before \begin{document} and uncomment 
%% following line to enable line numbers
%% \linenumbers

%% main text
%%

\section{Introduction}

In an environment where contractual relationships are becoming increasingly sophisticated, the challenges of transparency, security, and efficiency are intensifying considerably. Blockchain technology emerges as a fundamental solution to these challenges. Its decentralized and immutable nature not only reduces costs related to intermediaries but also minimizes fraud risks and improves compatibility between different systems. 

Among the major advances related to blockchain, smart contracts offer a unique perspective for automating and securing contractual clauses. Initially theorized by Nick Szabo \cite{szabo_smart_1996}, these autonomous programs guarantee strict compliance with conditions previously established between parties, without requiring human intervention. The advent of the Ethereum blockchain \cite{buterin_ethereum_2014} has enabled their large-scale deployment, benefiting from an optimized infrastructure for their decentralized development and execution. This evolution has promoted their adoption in various fields, from finance to supply chains. 

Despite their numerous advantages, smart contracts remain inaccessible to non-specialists, mainly due to their technical complexity \cite{frantz_institutions_2016}. Their implementation requires advanced programming skills and specific execution environments, constituting a major obstacle to their widespread adoption. To overcome these barriers, it is essential to propose more intuitive approaches, adapted to professionals wishing to effectively utilize these tools without immersing themselves in the underlying technical subtleties. 

Our multi-level model allows real-time monitoring of contractual execution and interactions via blockchain technology. This system ensures total transparency on the contract's evolution, thus facilitating transaction management and analysis of associated risks. Moreover, thanks to its multi-level architecture, it guarantees optimized performance, allowing rapid and instantaneous processing of contractual operations. Additionally, the multi-level architecture helps break down contractual complexity.

This study therefore aims to present a multi-level finite state machine model to represent and track the execution of smart contracts. This model allows a detailed representation of their structure while facilitating the evaluation of their functional and security properties. By proposing a formalized framework for the development of smart contracts, it aspires to become a valuable tool for professionals wishing to fully exploit these technologies without delving into the underlying technical complexities.

In this paper, we explore the potential of a Multi-Level Finite State Machine model to simplify smart contract development on blockchain technology. The following sections delve into the foundational aspects and innovative approaches that underpin our proposed solution. Section 2 reviews related work, examining existing methodologies and tools that have paved the way for smart contract automation and security. Section 3 elaborates on the multi-level representation, explaining how this hierarchical structure enhances contract modularity and traceability. Section 4 details the smart contract generation process, highlighting the use of reusable components and the concept of a package to ensure modularity. Section 5 discusses the security analysis of the model using state of the art tools. Finally, Section 6 concludes the study and outlines future perspectives, including the development of a Visual Programming Language (VPL) tool to further democratize smart contract creation.

\section{Related Work}

Smart contracts, positioned at the intersection of legal and technical challenges, are transforming contractual practices. Their recognition within existing legal frameworks remains an open question, while their automation relies on various innovative approaches that must overcome challenges related to security, complexity, and interoperability. First conceptualized by Nick Szabo \cite{szabo_smart_1996}, smart contracts are computer protocols that automate the execution of contractual terms. With the rise of blockchain technology introduced by Nakamoto in 2008 \cite{nakamoto_bitcoin_2009}, Ethereum \cite{buterin_ethereum_2014} provided a more flexible infrastructure through the Ethereum Virtual Machine, enabling the execution of complex smart contracts and fostering the development of decentralized finance and decentralized autonomous organizations. 

The automation of smart contract generation relies primarily on three main approaches. Model-driven approaches (MD) use modeling languages such as Unified Modeling Language (UML) \cite{garamvolgyi_model-2018}\cite{syahputra_uml_2019} or Business Process Model and Notation (BPMN) \cite{pintando_caterpillar_2018}\cite{tran_lorikeet_2018} to abstractly represent the structure and behavior of smart contracts. These models are then translated into executable code, as seen with Caterpillar \cite{pintando_caterpillar_2018}, which converts BPMN processes into Solidity code. However, this transformation remains a challenge due to the complexity of conversions and the constraints specific to blockchains. Domain-Specific Languages (DSLs) \cite{tateishi_dsl4sc_2019}\cite{skotnica_vpl_2020} offer a more tailored abstraction for smart contracts by using syntax and concepts specific to the contractual domain. Some, like DSL4SC \cite{tateishi_dsl4sc_2019}, generate state diagrams from formal models, while others, such as Ergo\footnote{Accord Project, Ergo. Available at: https://accordproject.org/projects/ergo/.}, focus on drafting legally executable contracts. Despite their flexibility, their adoption may be hindered by their dependence on specific platforms and a lack of standardization. Finally, Visual Programming Languages (VPLs) allow users to design smart contracts through graphical interfaces, making their development more accessible to non-programmers. VPL tools like Symboleo \cite{sharifi_symboleo_2020} or VeriSolid \cite{mavridou_verisolid_2019}, based on Finite State Machines, simplify the design and generation of Solidity code. However, these approaches still need to evolve to enhance security and ensure better blockchain interoperability.

Each approach presents strengths and limitations: MD facilitates conceptual representation but struggles to generate optimized code, DSLs provide precise formalization but are often confined to a specific ecosystem, and VPLs democratize development while requiring improvements to meet security and abstraction requirements. The evolution of these tools is essential to ensuring the robust and accessible integration of smart contracts into both technical and legal frameworks.
The proposed approach builds upon advanced modeling concepts, as introduced by the works of Suvorov and Ulyantsev \cite{suvorov_fsm_2019}, VeriSolid \cite{mavridou_verisolid_2019}, and Symboleo \cite{sharifi_symboleo_2020}, while addressing the limitations of these works by integrating a distributed and modular architecture through the concept multi-level finite state machine.

\newpage

\section{Multi-Level Representation}

Smart contract modeling relies on rigorous structuring of the various clauses and conditions to ensure reliable, automated execution. Among the most suitable approaches, the use of multi-level finite state machines offers a formal framework for managing both modularity and complex interactions between the different parts of the contract.

\begin{figure}[!htbp]
    \centering
    \includegraphics[width=1.0\textwidth]{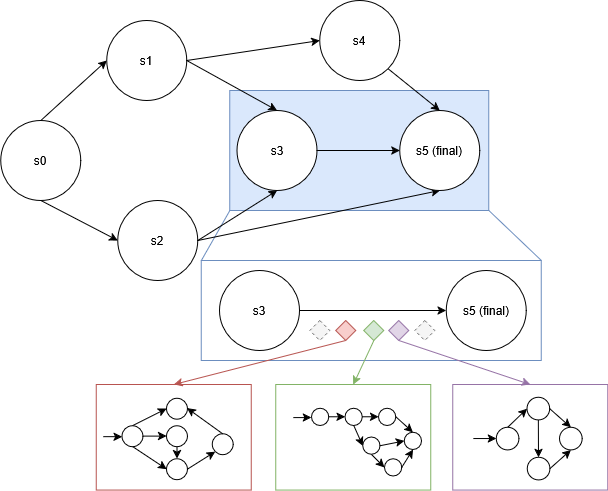}
    \caption{Multi-level representation.}\label{fig:msfsm-clauses}
\end{figure}

Unlike classical finite state machines, which operate on a single level of abstraction, the multi-level approach is based on a hierarchy of interconnected finite state machines. At each level, a finite state machine controls the transitions of a global process, while sub-machines detail more specific processes. This structure allows for dynamic dependencies between contract clauses, ensuring that changes in one state influence other aspects of the contract. 

The hierarchical structure of the multi-level finite state machine model provides several advantages:
\begin{itemize}
    \item Modularity: By breaking down the contract into multiple levels of abstraction, each level can focus on a specific aspect of the contract. This modularity simplifies the design and maintenance of complex contracts, as each module can be developed, tested, and updated independently.
    \item Traceability: The hierarchical structure enhances traceability by providing a clear and organized view of the contract's execution flow. This makes it easier to track the state of the contract and understand the interactions between different clauses.
    \item Flexibility: The multi-level approach allows for greater flexibility in contract design. Different levels can be used to represent different aspects of the contract, such as high-level business logic, intermediate processes, and low-level implementation details.
\end{itemize}

The multi-level finite state machine model also facilitates the management of dynamic dependencies between contract clauses. This is achieved through the use of interconnected state machines, where the state transitions in one machine can trigger transitions in other machines. This dynamic interaction ensures that the contract can adapt to changing conditions and requirements, providing a robust framework for managing complex contractual relationships.

Figure \ref{fig:msfsm-clauses} depicts the interactions among the clauses of a contract. Each clause is represented as an automaton, comprising states and transitions. The states correspond to the various steps within a clause, while the transitions facilitate progression from one step to another. These transitions are governed by safeguards, which we refer to as conditions. These conditions can result from the completion of other automata. We will explore later that these conditions may also be of different types.

The multi-level representation allows for a detailed and structured view of the contract's execution flow. Each clause is represented as an automaton, with states and transitions that define the steps and conditions for execution. The use of safeguards and conditions ensures that the contract can adapt to changing conditions and requirements, providing a robust framework for managing complex contractual relationships.

\section{Smart Contract Generation Process}

The process of generating smart contracts involves transforming abstract contractual models into deployable code, ensuring both functionality and security. This section outlines the methodology behind our approach, emphasizing the use of reusable components and a structured generation engine to streamline development and enhance reliability.

\begin{figure}[!ht]
    \centering
    \includegraphics[width=0.8\textwidth]{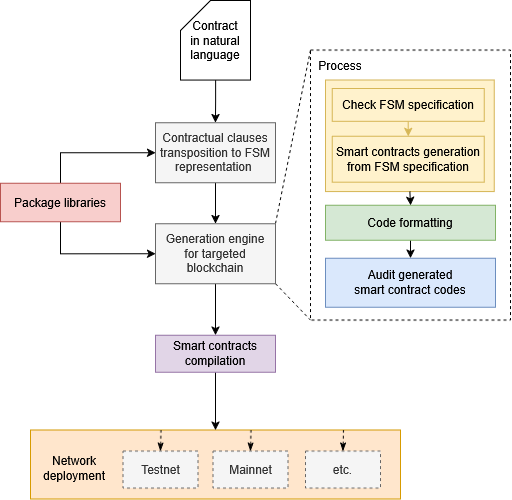}
    \caption{Generation and deployment processes.}\label{fig:msfsm-technical}
\end{figure}

Figure \ref{fig:msfsm-technical} illustrates the process of generating, compiling, and deploying a smart contract from a formal representation based on a finite state machine. This process consists of interconnected steps that progressively transform a contractual model into a deployable contract on a blockchain. Initially, the contract is described by clauses in natural language before being translated into a FSM representation. In the following sections, we will go through each step of this process in detail.

\subsection{Package Libraries}

Package libraries play a crucial role in our model. They contain essential reusable components for the generation of smart contracts. These libraries are used both in the construction of clause automata and during code generation. A package consists of functions, variables, and structures pre-developed by computer scientists, thus facilitating reuse and modularity. The format of these packages is JSON, which allows for easy integration and manipulation of the components within the generation engine. Notably, the functions within these packages can be utilized as conditions to facilitate transitions between states. This represents another type of condition mentioned earlier, where specific functions must be satisfied or executed to enable a transition from one state to another within the FSM representation.

\subsection{Contractual Clauses Transposition to FSM Representation}

Contractual clauses are currently manually transposed into a finite state machine (FSM) representation. This step allows structuring the clauses into states and transitions, thereby facilitating a more efficient and systematic management of contractual clauses. Although this transposition is currently performed manually, efforts are underway to automate this process to improve efficiency and precision. 

\begin{figure}[!ht]
    \centering
    \includegraphics[width=0.7\textwidth]{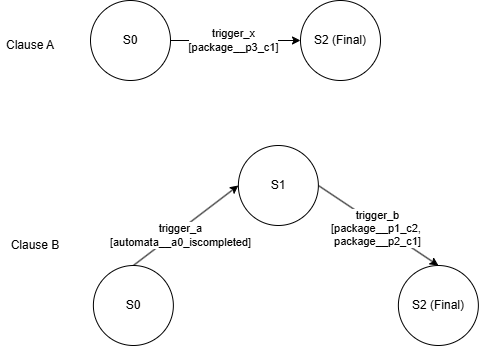}
    \caption{Simple clauses representation.}\label{fig:msfsm-simple}
\end{figure}

\newpage

\begin{lstlisting}[language=json,firstnumber=1,caption={FSM specification of Figure \ref{fig:msfsm-simple} clauses.},captionpos=b,label={lst:msfsm-simple}]
{
    "Clause A": {
        "states": [
            "s0",
            "s2"
        ],
        "transitions": [
            {
                "source": "s0",
                "destination": "s2",
                "trigger": "trigger_x",
                "conditions": [
                    "package__p3_c1"
                ]
            },
        ]
    },
    "Clause B": {
        "states": [
            "s0",
            "s1",
            "s2"
        ],
        "transitions": [
            {
                "source": "s0",
                "destination": "s2",
                "trigger": "trigger_a",
                "conditions": [
                    "automata__a0_iscompleted"
                ]
            },
            {
                "source": "s1",
                "destination": "s2",
                "trigger": "trigger_b",
                "conditions": [
                    "package__p1_c2"
                    "package__p2_c1"
                ]
            }
        ]
    },
}
\end{lstlisting}

Figures \ref{fig:msfsm-simple} and the specification \ref{lst:msfsm-simple} primarily illustrate the results of the process of transposing natural language, as used in the contract, into an multi-level finite state machine specification. Currently, this transposition is performed manually; however, it is conceivable to design an automated system to facilitate this conversion. The multi-level specification encompasses all states, as well as transitions defined by their source and destination states, triggers, and associated conditions. These elements constitute the various operational clauses of the contract, thereby enabling precise and structured modeling of the described processes.

\subsection{Generation Engine and Smart Contract Compilation}

We have developed a generation engine specifically designed to create smart contracts tailored to a target blockchain. Initially, we have utilized the Ethereum blockchain for this purpose, employing tools such as py-solc-x\footnote{py-solc-x. Available at: https://solcx.readthedocs.io/en/latest/.} for compilation and web3.py\footnote{web3.py. Available at: https://web3py.readthedocs.io/en/stable/.} for deployment and execution of smart contracts. This engine utilizes the FSM representations of clauses to generate appropriate code, ensuring that the contracts are adapted to the specificities of the targeted blockchain.

% détailler le tri topologique, construction de l'arbre. Retirer l'ambiguïté du contrat

\begin{itemize}
    \item FSM Specification Verification: This step involves verifying the FSM specification to ensure that all clauses are correctly defined and structured.
    \item Smart Contract Generation from FSM Specification: This phase involves converting states and transitions into executable code, thereby enabling the creation of functional smart contracts. The contracts are generated in a topological order because some depend on other smart contracts, which can potentially create cycles.
    \item Code Formatting: The generated code is formatted to ensure it is readable and compliant with coding standards, guaranteeing optimal quality and maintainability.
    \item Audit of Generated Smart Contract Codes: A code audit is conducted to verify security, robustness, and compliance with contractual requirements, ensuring the reliability of the deployed contracts. In the context of utilizing the Ethereum blockchain, the framework Smartbugs \cite{joao_smartbugs_2023}. We will discuss later in the Security Analysis section.
\end{itemize}

\subsection{Deployment}

The deployment phase is essential to ensure that smart contracts function as intended and are ready for production use. Compiled smart contracts are first deployed on the blockchain, a crucial step that allows them to be implemented in a real environment where they can be executed and interact with other contracts and users. It is notably possible to execute and monitor the various clauses of the contract during this phase. Next, the contracts are deployed and tested on a test network, or Testnet, to ensure they function as expected. The Testnet is a test environment that replicates the conditions of the main network but without the risks associated with using real assets or sensitive data, thus allowing for the detection and correction of any potential bugs or vulnerabilities before final deployment. After validation on the Testnet, the contracts are finally deployed on the main network, or Mainnet, marking their transition to production where they are accessible to end-users and interact with real assets. Deployment on the Mainnet is a critical step that requires meticulous preparation to avoid any malfunctions or security flaws.

\section{Security analysis}

In our security analysis, we utilized the SmartBugs framework \cite{joao_smartbugs_2023}. This framework is particularly useful as it consolidates 20 different security tools, including well-known tools such as Slither\footnote{Slither. Available at: \textit{https://github.com/crytic/slither}.}, Solhint\footnote{Solhint. Available at: \textit{https://github.com/protofire/solhint}.}, and Mythril\footnote{Mythril. Available at: \textit{https://github.com/ConsenSysDiligence/mythril}.}. The advantage of using SmartBugs is that it significantly simplifies the process of configuring and running these tools, which can often be time-consuming and complex.

\begin{figure}[!ht]
    \centering
    \includegraphics[width=0.8\textwidth]{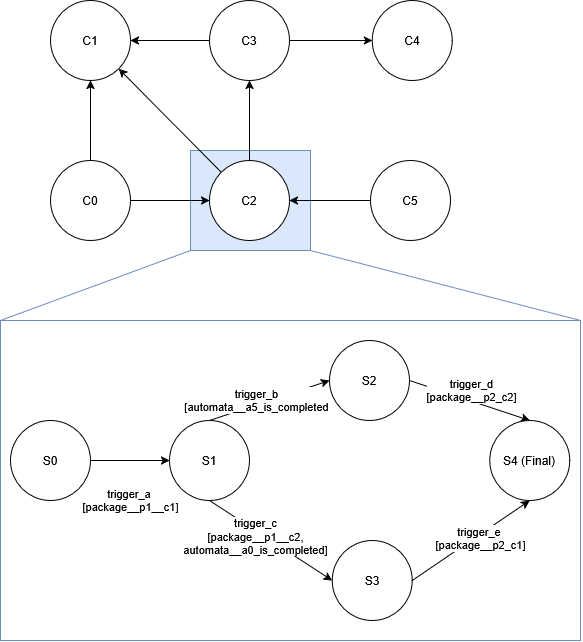}
    \caption{Dependency graph of test contract with clause 2 detailed.}\label{fig:dependency-graph}
\end{figure}

To evaluate potential vulnerabilities in our model, we generated test smart contracts using our multi-level finite state machine modeling approach. These contracts were then subjected to rigorous testing using the SmartBugs framework with the Solidity compiler version 0.8.29 to identify any security flaws. to identify any security flaws. The figure \ref{fig:dependency-graph} illustrates the dependency graph of the contract clauses. It is provided for indicative purposes. Each node within the graph represents a specific clause of the contract, while the transitions between these nodes signify the dependencies among the clauses. It is important to note that cyclic dependencies between clauses are impossible due to the implementation of a topological sort. This figure also provides a detailed view of Clause 2, which depends on Clause 0 and Clause 5. It demonstrates the use of conditions derived from packages and the utilization of functions originating from other automata within the test contract.

We focused on evaluating the primary vulnerabilities commonly associated with smart contracts, as identified by various researchers \cite{he_security_2024}\cite{li_survey_2020}\cite{atzei_survey_2017}\cite{zhu_survey_2020}. These vulnerabilities include reentrancy, integer overflow, unchecked low-level calls, Tx.origin issues, time manipulation, delegate call, access control, Transaction-Ordering Dependence, and denial of service. Our analysis revealed that none of these vulnerabilities were present in the generated smart contracts. It is important to note that prior to using the generator, all functions, variables, and structures sourced from established packages were thoroughly audited to ensure their integrity and security.

\section{Conclusion and Perspectives}

In this study, we presented a multi-level finite state machine model that significantly simplifies the development of smart contracts on blockchain technology. By leveraging reusable components and a structured generation engine, our approach offers a robust and accessible solution for professionals wishing to fully exploit these technologies without delving into the underlying technical details. The hierarchical structure of our model enhances the modularity and traceability of contracts, thus facilitating their management and updates.

Our major contribution lies in the ability to automate and certify the proper execution of contracts while reducing the technical complexity for end-users. This not only paves the way for wider adoption of smart contracts across various sectors, from finance to supply chain management, but also makes these technologies more accessible and reliable by providing a clearer and more structured approach to contract development.

Looking ahead, there are numerous opportunities for further enhancement. Developing a Visual Programming Language (VPL) tool could make smart contract creation even more accessible, opening up these technologies to an even broader audience. Applying our model to real-world scenarios, such as supply chain management, would not only validate its effectiveness but also highlight areas for potential improvements. Additionally, integrating automated testing frameworks could further enhance the robustness of the contracts, ensuring their reliability in diverse applications. Exploring compatibility with various blockchains could also broaden the applicability of our model, making it more versatile and adaptable to different environments.

In conclusion, our model provides a solid and innovative framework for smart contract development, significantly reducing complexity and enhancing clarity. The future perspectives we have outlined offer exciting opportunities to further enhance its capabilities and applicability, ultimately contributing to the broader adoption and success of smart contracts in various industries.

\section*{Acknowledgements}

This work was funded by the Region of Normandy, Le Havre Seine Métropole, and the Agence Nationale de la Recherche (ANR) as part of the SmartLogiLab project in the labcom program (ANR-22-LCV2-0013) which is a collaboration between SOGET and LITIS.

\end{document}